# High Critical Current Density and Improved Irreversibility Field in Bulk MgB$_2$ Made By A Scaleable, Nanoparticle Addition Route

J. Wang, Y. Bugoslavsky, A. Berenov, L. Cowey, A.D. Caplin, L.F. Cohen, J.L. MacManus Driscoll
*Centre for High Temperature Superconductivity, Imperial College, Prince Consort Rd., London SW7 2AZ, UK.*

L. D. Cooley, X. Song, D. C. Larbalestier
*Applied Superconductivity Center, University of Wisconsin-Madison, 1500 Engineering Dr., Madison, WI, 53706-1687, U.S.A.*

## Abstract

Bulk samples of MgB$_2$ were prepared with 5, 10, and 15% wt.% Y$_2$O$_3$ nanoparticles added using a simple solid-state reaction route. Transmission electron microscopy (TEM) showed a fine nanostructure consisting of ~3-5 nm YB$_4$ nanoparticles embedded within MgB$_2$ grains of ~400 nm size. Compared to an undoped control sample, an improvement in the in-field critical current density $J_C$ was observed, most notably for 10% doping. At 4.2K, the lower bound $J_C$ value was ~$2 \times 10^5$ A.cm$^{-2}$ at 2T. At 20K, the corresponding value was ~$8 \times 10^4$ A.cm$^{-2}$. Irreversibility fields were 11.5T at 4.2K and 5.5T at 20K, compared to ~4T and ~8T, respectively, for high-pressure synthesized bulk samples.

In slightly more than one year after the discovery of superconductivity in magnesium diboride, there is now a wide body of evidence indicating that $MgB_2$ does not contain intrinsic obstacles to current flow between grains, unlike the high-temperature superconducting cuprates. Evidence for strongly coupled grains has been found even in randomly aligned, porous, and impure samples [1, 2], suggesting that dense forms of $MgB_2$ will be attractive in high-current applications at 20-30 K and perhaps 4.2 K. So far, however, bulk samples have demonstrated modest values of the irreversibility field $\mu_0H^* (T)$ reaching about 4 T at 20 K and 8T at 4.2 K [3]. For comparison, established low temperature superconductors, e.g. NbTi (10 T) and $Nb_3Sn$ (20 T), have significantly higher irreversibility fields at 4.2K, while $Bi_2Sr_2Ca_2Cu_3O_{10}$ (3 T) is becoming established at 20K [4]. $MgB_2$ tape results are somewhat more promising, with $\mu_0H^*$ values of above 5 at 20 K [5, 6, 7, 8], where partial orientation of crystallites parallel to the field is playing a role. Since the irreversibility field is the practical limit to magnet applications, it is desirable to make $\mu_0H^*$ values as high as possible.

A central question is how to further increase the irreversibility field in addition to introducing crystallographic texture. Alloying additions, such as atomic substitution for Mg or B or added interstitial atoms, increase electron scattering and decrease the coherence length, producing higher upper critical and irreversibility fields [9, 10]. Adding nanometer-scale defects can produce similar effects. For example, proton irradiation studies showed that $\mu_0 H^*$ increased significantly from ~3.5 T to ~ 6 T at 20 K with only moderate damage, corresponding to atomic displacements of a few %, due to either vacancies or interstitials [11]. Mechanical processing also produces structural defects, and similar increases in the irreversibility field have been reported [6, 8, 12]. These increases were steeper than the concomitant reductions in the critical temperature $T_c$, suggesting it is viable to improve the accessible field range without sacrificing other superconducting properties too much. So far, however, it has been difficult to separate the effects chemical changes from structural changes, since both were present in the experiments above.

To explore more practical and scaleable routes to defect incorporation in bulk $MgB_2$, the present study explores chemical and nanostructural changes via addition of nanoparticles. While MgO seems the most suitable second phase nanoparticle, we chose $Y_2O_3$ nanoparticles owing to the fact that it can be purchased cheaply and in large quantities. Cimberle et al. [13] found that $J_C$ increased by up to a factor of 3 for Li-, Al- and Si- doped samples, although $\mu_0 H^*(T)$ remained unchanged at ~ 4 T for 20 K. Likewise, Feng et al. [14, 15] claimed much higher $J_C$ for Ti and Zr doped samples, reaching 5 x $10^5$ A/cm$^2$ at 1 T and 20 K although less dramatic differences could be seen for $\mu_0 H^*$, which was ~4.0T at 20 K

In the present Letter, we show that by the nanoparticle addition $\mu_0 H^*$ is increased to 5.5T at 20 K and 11.5 T at 4.2 K, with an accompanying increase of $J_c$ to ~$10^5$ A/cm$^2$ at 2 T and 20 K. A key finding, obtained by high-resolution transmission microscopy and spectroscopy, is evidence for a regular distribution of $YB_4$ nanoprecipitates, formed by reaction of the $Y_2O_3$ nanoparticles with B. This suggests other oxide nanoparticles that also form stable borides will give a similar nanostructure and convey similar benefits to flux pinning and $\mu_0 H^*$.

Doped samples were prepared from amorphous boron powder (99%, Fluka), 5, 10, or 15 wt.% $Y_2O_3$ nanoparticles (15-30 nm particle size, 99.5%, Pi-Kem), and Mg powder (99%, Riedel- de Haen) by mixing and pressing into 5 mm diameter by 2 mm thick pellets. Undoped $MgB_2$ pellets were similarly prepared to the control samples. Reactions were performed at 900°C in a reducing gas mixture of Ar-2%$H_2$. In order to counteract the effects of Mg loss, Mg foil was present in the reaction vessel during the reaction. The heating and cooling rates used were ~20 °C/min, and the dwell time at the peak temperature was 15 min.

The samples were clearly macroscopically porous when viewed by light microscopy (not shown). The geometrical densities for the different pellets were measured to be 50±5%.

Figure 1a shows the results of XRD analyses for the series of doped samples compared to an undoped sample. In addition to $MgB_2$, small quantities of MgO and, in the doped samples, $YB_4$ are indicated; there are no peaks corresponding to either pure Mg or $Y_2O_3$. Hence, in the doped samples, it can be concluded that the $Y_2O_3$ reacted with B to form $YB_4$. Since this decreased the amount of B available for reaction to $MgB_2$, excess Mg either was transported away from the pellet or became oxidized. Indeed, an increase of the MgO peak intensities correlates with increasing $Y_2O_3$ fraction in Fig. 1a. The volume of the $MgB_2$ unit cell was calculated as a function of $Y_2O_3$ fraction, which is shown in Figure 1b. The UnitCell program was used to refine the lattice parameters and calculate second phase particle sizes. There is possibly only a slight change in the unit cell volume with increased doping. This change might represent incorporation of oxygen into the lattice, or it could be due to strain from the added nanoparticles.

Fig. 2a shows a TEM diffraction contrast image of the 10% $Y_2O_3$ sample. The grain size of the $MgB_2$ is ~400 nm and precipitates at two different levels are seen. Precipitates with ~10 nm size occur at the $MgB_2$ grain boundaries (region 1), while inside the $MgB_2$ grain interior (region 2), evenly distributed, 3 to 5 nm precipitates are seen. A magnification of region 2 is shown in the inset of Fig. 2a. Selected area diffraction patterns taken from both regions were very similar. Fig. 2b shows the diffraction pattern along the $MgB_2$ [120] direction. The circled spots are consistent with $MgB_2$ and the indexed rings with $YB_4$.

Although all diffraction rings from MgO are contained in the more complex $YB_4$ pattern, additional rings were present for $YB_4$ which could not be due to MgO and therefore confirm that most of the precipitates are $YB_4$. Large regions of MgO were observed in different areas of the sample and were found with 40 to 200 nm size. Diffraction patterns confirm the MgO structure, however $YB_4$ precipitates are also present within the MgO regions. The maximum particle sizes calculated from XRD for $YB_4$ and MgO were ~10 nm and ~150 nm, respectively, consistent with the TEM analyses.

Critical temperature $T_c$ values were obtained by measuring the magnetic moment versus temperature $m(T)$ using a vibrating sample magnetometer (VSM), shown in Fig. 3. Samples were zero-field cooled and then warmed from 10 K in an applied field of 5 mT. Similar transitions, with an onset at ~39 K and an endpoint at ~38 K, are seen for the control, 5% and 10% $Y_2O_3$ samples, but these values are reduced by ~1 K for the 15% sample.

Fig. 4 shows $J_C(H)$ at 20 K for the series of doped $MgB_2$ samples, as well as the undoped sample, a fragment of a sample from a commercial source, the 10 at.% Zr-doped sample of Feng et al.[14] and high-pressure synthesized $MgB_2$ [16]. The inset of Fig. 4 shows $J_C(H)$ at 4.2K for the 10 wt. % $Y_2O_3$ doped sample. All samples were measured in a VSM and the Bean model was used to deduce the critical current density from the magnetization hysteresis [17]. For our measurements, in fields of < 1 T the apparent plateau in $J_C$ is artificial, due to saturation of the magnetometer. Therefore, the actual values are higher than shown. Our $J_C$ values are based on full sample connectivity and are multiplied by 2 to allow for the 50% porosity. In fact, associated with the porosity is a reduction in the grain-to-grain contact area, and so a restriction of the cross-section available for current flow. Also, the grain boundaries in the doped samples are partially obstructed by precipitates (Fig. 2a).

Consequently these estimates are only lower bounds. Nonetheless, $J_C$ is high and comparable to the fully dense, high pressure synthesized material. The doping level of 10% produced the largest difference in both $J_C$ and $\mu_0 H^*$. At 2T, $J_C$ for the sample is ~8 x $10^4$ A cm$^{-2}$, a factor of ~3 higher than for the undoped pellet sample and ~4 higher than the fragment. Compared to the Zr-doped sample, our low-field $J_C$ values are lower in the ~1-2 T region, but higher thereafter.

The $\mu_0 H^*$ value for the 10% $Y_2O_3$ doped sample, as defined by a critical current density criterion of $10^2$ A/cm$^2$, was 11.5T at 4.2K and 5.5T at 20K. The value at 20K was confirmed to be 5.5T from creep rate measurements (not shown), and is ~1.5 times higher than both the high-pressure and Zr-doped samples. The 4.2K value is comparable with that for Nb-Ti. It is interesting that the control sample also has an enhanced irreversibility field, and this point is discussed further below.

The rapid formation technique used in this experiment apparently produced different superconducting properties, relative to those of high-pressure synthesized bulk, in all of the samples. However, at <4T a higher irreversibility field and higher critical current density are seen for the doped samples, suggesting the additional effects of the nanoparticles. At >5T, the undoped sample outperforms the doped samples, but this is most likely related to the greater connectivity of grain boundaries in the undoped sample, the doped samples having additional phases at the grain boundaries.

Interestingly, the samples show similar $J_C$ and $\mu_0 H^*$ values to a recent report describing heavily ball milled, nanocrystalline powders of Gumbel et al.[12]. However, ball milling also reduced $T_c$ to 34.5K, suggesting that disorder or possibly alloying (from the milling process) strongly affects superconductivity in $MgB_2$. Reduced $T_c$'s in the Zr-doped samples is also indicative of alloying [14]. In the present work, the critical temperature remains near 39 K, and there is a weak increase (if any) in the unit cell volume. We believe, therefore, that the nanoparticle additions neither alloyed the surrounding $MgB_2$, nor produced significant disorder. The observed increase in $\mu_0 H^*$ may be due to increasing the number of point scattering sites, since the observed precipitates are 3-5 nm in size and uniformly distributed within the grains.

**Summary and Conclusions**

In summary, we have shown that incorporating $Y_2O_3$ nanoparticles together with Mg and B powders results in the formation of $MgB_2$ with a uniform dispersion of $YB_4$ nanoprecipitates. This nanostructure was achieved using a reaction at 900 °C for 15 minutes. The precipitates have 3 to 5 nm size, with larger ~10 nm precipitates occurring at some grain boundaries. We find little if any increase in the $MgB_2$ unit cell and no change in the critical temperature, suggesting that neither alloying nor strong disorder accompanied the reaction. At 20K, the critical current density deduced by magnetization is $>10^5$ A/cm$^2$ in low fields, comparable to that of high-pressure synthesized bulk. Significant increases in the irreversibility field were also observed, of 11.5T at 4.2K and 5.5 T at 20 K. The best properties were obtained by doping with 10wt.% $Y_2O_3$.

**Acknowledgements**

The work at Imperial College was supported by EPSRC and work at Wisconsin by the US Department of Energy and the National Science Foundation.

**Figure Captions**

**Figure 1:** x-ray diffraction of $Y_2O_3$-doped $MgB_2$. a) diffractograms of undoped, 5-, 10- and 15 wt. % $Y_2O_3$-doped $MgB_2$, and b) volume of $MgB_2$ cell versus wt. % $Y_2O_3$.

**Figure 2:** TEM of 10 wt. % $Y_2O_3$ doped $MgB_2$, a) micrograph showing nanoprecipitates of $YB_4$ embedded in $MgB_2$ grains, and b) diffraction pattern along $MgB_2$ [120] direction. $YB_4$ ring pattern is outlined.

**Figure 3:** Normalised d.c. magnetic susceptibility versus temperature for doped and undoped pellet samples. The demagnetisation factors, n, were evaluated using the external sample dimensions.

**Figure 4:** $J_C$ *(H)* at 20 K for the series of $Y_2O_3$ doped $MgB_2$ pellet samples, as well as an undoped pellet, a fragment of a sample from a commercial source (Alfa Aesar). For comparison, the 10 at.% Zr-doped sample of Feng et al.[14] and high-pressure synthesized $MgB_2$ [18] are included. The inset shows $J_C$ *(H)* at 4.2K for the 10 wt. % $Y_2O_3$ doped sample.

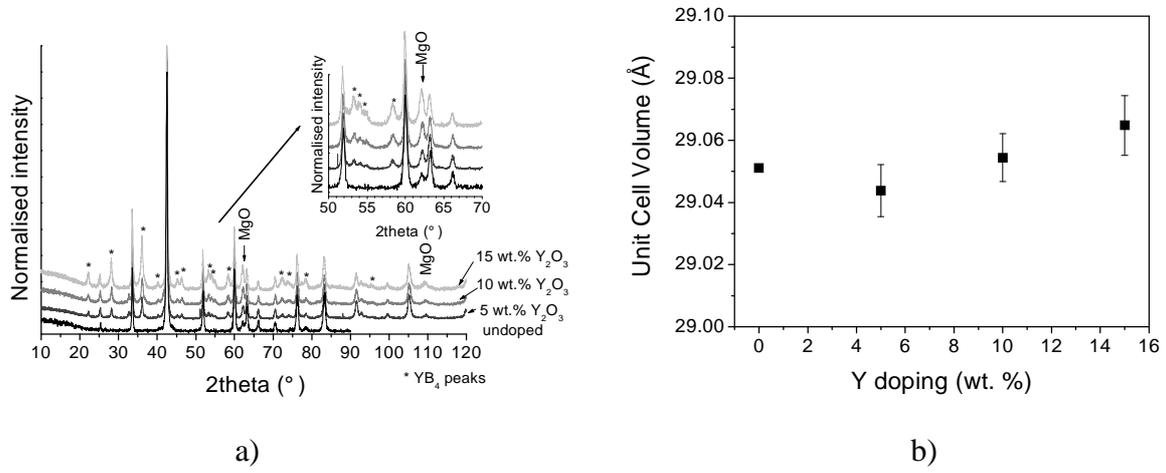

Figure 1

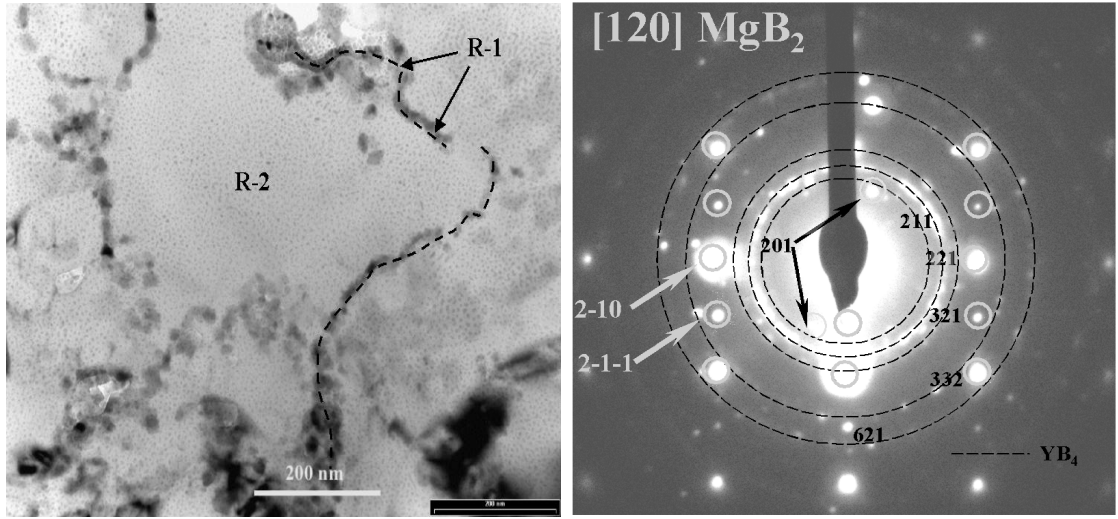

a)                                                      b)

Figure 2

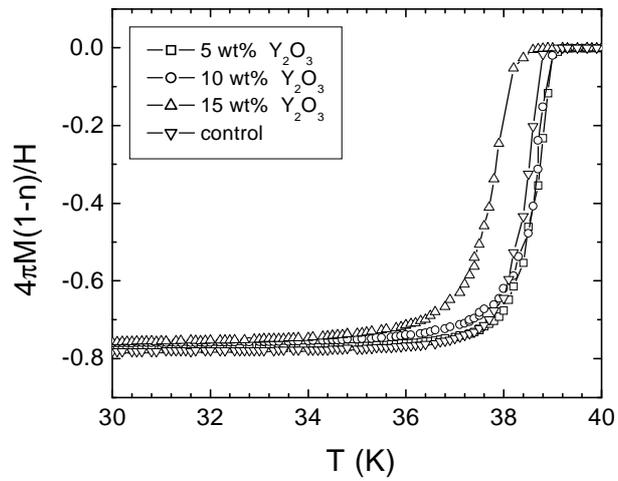

Figure 3

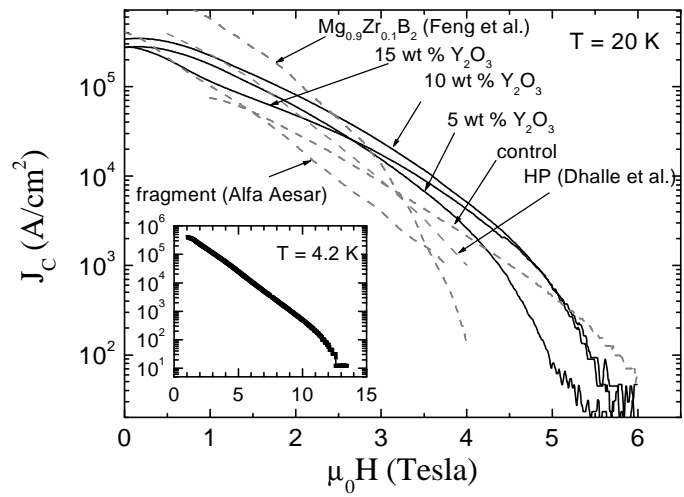

Figure 4